\documentclass[aps,prb,article,twocolumn,showpacs,preprintnumbers,amsmath,amssymb,superscriptaddress]{revtex4}

\date{\today}
\usepackage{epsfig}
\usepackage{subfigure}
\usepackage{graphicx}
\usepackage{dcolumn}
\usepackage{bm}
\usepackage[colorlinks,linkcolor=blue,hyperindex,CJKbookmarks]{hyperref}
\usepackage{float}
\usepackage{hyperref}
\usepackage{comment}
\setcitestyle{numbers,square}

\newcommand{\Ham}   {{\mathcal{H}}}

\newcommand{\Schrdg} {{Schr\"{o}dinger}}
\begin{document}

\title{Producing Coherent Excitations in Pumped Mott Antiferromagnetic Insulators}
\author{Yao Wang }
 \affiliation{Department of Applied Physics, Stanford University, California 94305, USA}
 \affiliation{Stanford Institute for Materials and Energy Sciences, SLAC National Accelerator Laboratory, 2575 Sand Hill Road,
Menlo Park, California 94025, USA}
\author{Martin Claassen}%
\affiliation{Department of Applied Physics, Stanford University, California 94305, USA}%
\affiliation{Stanford Institute for Materials and Energy Sciences, SLAC National Accelerator Laboratory, 2575 Sand Hill Road,
Menlo Park, California 94025, USA}
\author{B. Moritz}
\affiliation{Stanford Institute for Materials and Energy Sciences, SLAC National Accelerator Laboratory, 2575 Sand Hill Road,
Menlo Park, California 94025, USA}%
\affiliation{Department of Physics and Astrophysics, University of North Dakota, Grand Forks, North Dakota 58202, USA}
\author{T. P. Devereaux}
 \email[Author to whom correspondence should be addressed to Y. W. (\href{mailto:yaowang@g.harvard.edu}{yaowang@g.harvard.edu}) or T.P. D. (\href{mailto:tpd@stanford.edu}{tpd@stanford.edu})
]{}
\affiliation{Stanford Institute for Materials and Energy Sciences, SLAC National Accelerator Laboratory, 2575 Sand Hill Road,
Menlo Park, California 94025, USA}%
\affiliation{Geballe Laboratory for Advanced Materials, Stanford University, California 94305, USA}
\date{\today}
\begin{abstract}
Nonequilibrium dynamics in correlated materials has attracted attention due to the possibility of characterizing, tuning, and creating complex ordered states. 
To understand the photoinduced microscopic dynamics, especially the linkage under realistic pump conditions between transient states and remnant elementary excitations, we performed nonperturbative simulations of various time-resolved spectroscopies. We used the Mott antiferromagnetic insulator as a model platform. 
The transient dynamics of multi-particle excitations can be attributed to the interplay between Floquet virtual states and a modification of the density of states, in which interactions induce a spectral weight transfer. 
Using an autocorrelation of the time-dependent spectral function, we show that resonance of the virtual states with the upper Hubbard band in the Mott insulator provides the route towards manipulating the electronic distribution and modifying charge and spin excitations. 
Our results link transient dynamics to the nature of many-body excitations and provide an opportunity to design nonequilibrium states of matter via tuned laser pulses.
\end{abstract}

\maketitle

\section{Introduction}
\begin{figure}[!t]
\begin{center}
\includegraphics[width=8.5cm]{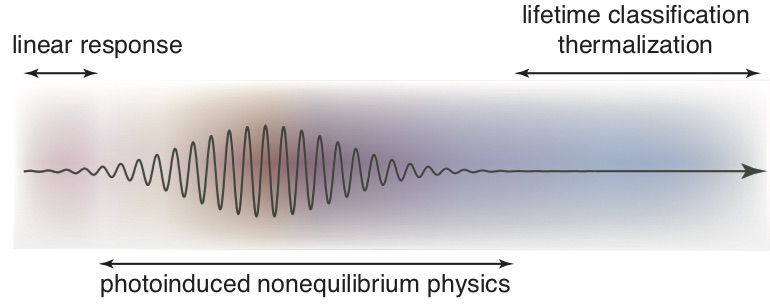}
\caption{\label{fig:0} Schematic cartoon illustrating various time regions during a pump process.
}
\end{center}
\end{figure}

Time-domain techniques have become increasingly useful over the past decade due to their potential for characterizing elementary excitations, manipulating intertwined orders, and creating new states of matter\cite{fausti2011light, lee2012phase, kim2012ultrafast, dal2014snapshots, wang2014real, gorshkov2011tunable, rubbo2011resonantly, carrasquilla2013scaling, hazzard2013far, schachenmayer2013entanglement}.
Ultrafast pump lasers are particularly powerful tools, with ground-breaking experiments demonstrating control of competing electronic orders and the associated elementary excitations\cite{rini2007control, Yonemitsu20081, PhysRevLett.102.106405, tomeljak2009dynamics, schmitt2008transient,  lu2012enhanced, PhysRevB.88.045107,rincon2014photoexcitation}.
Various regimes in a pump-probe experiment contain information about different physical processes, as shown in Fig.~\ref{fig:0}.
The long-time recovery following a pump can be used to quantify lifetimes and classify interaction mechanisms\cite{dal2012disentangling, hellmann2012time, yang2015inequivalence}, while the weak pre-pump tail falls in the linear response regime\cite{reed2010effective, wang2014real}.
Between these two limits, the nonequilibrium dynamics and related photomanipulation during or shortly after the pump also contains rich information about the underlying physics\cite{wang2013observation, mahmood2016selective, claassen2017dynamical}.
Due to the complexity of this regime, there are usually two extreme scenarios to simplify the problem: 
Floquet theory, which assumes an infinitely long, periodic pump\cite{Floquettheorem, Shirley:1965cy, Barone:1977ur, moskalets2002Floquet, dehghani2014dissipative, sie2015valley, Sentef:2015jp}; and an instantaneous quantum quench, which involves a discontinuous change of parameters\cite{fausti2011light, schmitt2008transient, wang2016using}.
However, the nonequilibrium behavior in different systems can deviate from either of these scenarios under a realistic pump condition. 

This problem becomes particularly important in strongly-correlated materials with various intertwined orders and emergent phase transitions. 
In such systems, the charge and spin excitations are significant due to their fundamental role in emergent phenomena\cite{kivelson2003detect, keimer2015quantum}.
Recently, Floquet studies have been extended into many-body physics, with a focus on the engineering of effective Hamiltonians obtained perturbatively to the leading order \cite{bukov2016schrieffer,bukov2016heating,bukov2015prethermal, kuwahara2016Floquet,eckardt2017colloquium}. 
However, these transient effective Hamiltonians fail to provide information about switch-on/off as well as evolving occupations during a pump with a realistic time profile\cite{mendoza2017ultra,coulthard2016enhancement}. 
In other words, the analysis based on discrete-time-translational invariance cannot correctly capture the transition between Floquet states connected to the remnant excitations. 
Such information is significant to the understanding of quasiparticle population dynamics and decay\cite{basov2017towards}. It also helps to characterize the impact on the collective modes or competing orders that can be manipulated by ultrafast techniques\cite{wang2017light}.
Therefore, to connect these concepts and understand or photocontrol the underlying physics requires a pure nonequilibrium study of the microscopic dynamics under realistic pump conditions.
This problem includes, but is not restricted to, interpreting how the coherent quasiparticles or collective excitations behave during realistic pumps and determining whether they can be selectively tuned by the pumped laser, an important question for designing nonequilibrium states of matter.

Specifically in a Mott insulator where rich emergent phenomena arise, ultrafast studies have been performed observing a suppression of the spin order.
Without an exact interpretation of the transient dynamics, this phenomenon was attributed to the melting of the Mott gap due to effective heating\cite{perfetti2008femtosecond}. 
This incoherent heating constitutes another extreme for the interpretation of an ultrafast process in gapped systems, in addition to the coherent quantum quench and Floquet physics.
Thus, a detailed pump-probe study beyond the single-particle level is required to dynamically correlate these scenarios at various time scales during a realistic pump and reveal the nature of elementary excitations.

To address these issues from a microscopic perspective, we perform a time-resolved exact-diagonalization study on the single-band Hubbard model. 
Starting from an insulator at half-filling, we focus on driving the system at pump frequencies close to resonance with the upper Hubbard band, thereby transiently suppressing antiferromagnetism. While preserving the original Mott gap, the time-resolved dynamical charge and spin spectra reflect a suppression of spin order and development of low-energy charge excitations through a series of Floquet-like spectral structures. 
By comparing the multi- and single-particle spectra out of equilibrium, the dynamics of these elementary excitations can be interpreted as stemming from the interplay between Floquet virtual states and interaction effects, modulated by the finite pump profile. 
A detailed analysis reveals that remnant excitations develop from resonance between Floquet virtual states and the upper Hubbard band, leading to coherent excitation as opposed to incoherent heating.

\begin{figure}[!t]
\begin{center}
\includegraphics[width=0.9\columnwidth]{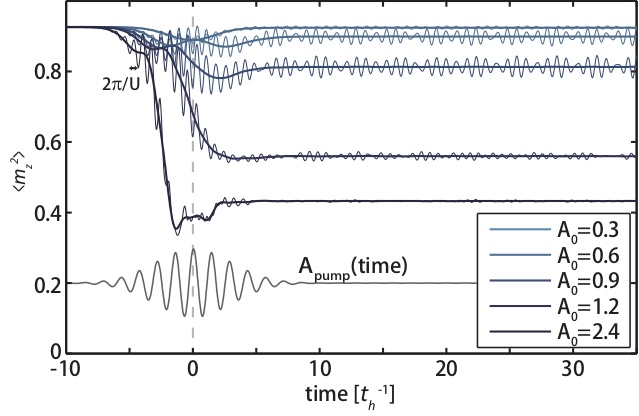}
\caption{\label{fig:1} (Color online) Instantaneous magnetic fluctuation $\langle\hat{m}_z^2\rangle(t)$. Various curves represent different pump intensities but fixed frequency ($\Omega = 4.4t_h$) and width ($\sigma_t = 3.0t_h^{-1}$). The phase-averaged magnetic moments are shown in the central thick curves. The gray curve represents the pump field.  
}
\end{center}
\end{figure}

\section{Nonequilibrium Dynamics}
We simulate the pulsed laser field by the time-dependent electric field (vector potential) along the $x$-direction
\begin{eqnarray}
A(t)=A_0 e^{-(t-t_0)^2/2\sigma_t^2}\cos[\Omega (t-t_0) + \phi].
\end{eqnarray}
Starting from a single-band one-dimensional Hubbard model of correlated electrons, the pump field enters via a Peierls' substitution. 
\begin{eqnarray}
\Ham(t)=-t_h \sum_{i\sigma} \left[ e^{iA(t)} c^\dagger_{i\sigma}c_{i+1\sigma}+h.c.\right]+U\sum_i n_{i\uparrow}n_{i\downarrow}.
\end{eqnarray}
Here, $c_{i\sigma}^\dagger$ ($c_{i\sigma}$) creates (annihilates) an electron at site $i$ with spin $\sigma$. The parameter $t_h$ is the nearest-neighbor hopping integral and $U$ is the on-site Coulomb repulsion.
We use the parallel Arnoldi method\cite{lehoucq1998arpack} to determine the ground state wavefunction\cite{jia2017paradeisos} and the Krylov subspace technique\cite{balzer2012krylov, park1986unitary, hochbruck1997krylov, moler2003nineteen} to evaluate the evolution of a state $|\psi(t\!+\!\delta t)\rangle\!=\!e^{-i\mathcal{H}(t)\delta t}|\psi(t)\rangle$. $U=8t_h$ is set to simulate strong correlations. We choose a chain size to be $L=12$ and a coarse-grained time set by $\delta t\!=\!0.05\,t_h^{-1}$. 

In equilibrium, the model exhibits spin-charge separation with gapless spinon and gapped charge excitations ($\Delta_c \sim U$)\cite{lieb1968absence, essler2005one}.
Because the pump field only couples to the charge degrees of freedom, the pumped spin and charge physics are expected to behave differently, inducing emergent excitations at various orders of magnitude.

We monitor the magnetic fluctuation $\hat{m}_z^2$ = $\sum_i \big(c_{i\uparrow}^\dagger c_{i\uparrow} - c_{i\downarrow}^\dagger c_{i\downarrow}\big)^2$ out of equilibrium.  
It displays a suppression accompanied by a fast oscillation with periodicity $\sim\!2\pi/U$ [shown for a fixed $\phi\!=\!0$ in Fig.~\ref{fig:1}], reflecting the fast scattering across the Mott gap.
Averaging over the pump phases as appropriate for most experimental measurements with finite time resolution, the magnetic moments show that a stronger pump suppresses the intrinsic magnetism due to photo-induced doublon generation, which may at first glance be attributed to an effective heating-induced Mott gap closure. 
However, neither spectra nor energy fitting can extract a consistent effective temperature capturing the basic post-pump features [see the discussions in Appendix~\ref{app:1}]. This indicates that the underlying physics cannot be mapped to an incoherent, quasi-equilibrium heating induced by pump fluences. [Note here we referred to a heating effect in an ultrafast process, rather than the long-term thermalization discussed in Ref.~\onlinecite{bukov2016heating}.]
In the following text, we show this is instead a nonequilibrium effect associated with coherent excitations.

\section{Pump-Probe Spectroscopies}
\begin{figure*}[!ht]
\begin{center}
\includegraphics[width=14cm]{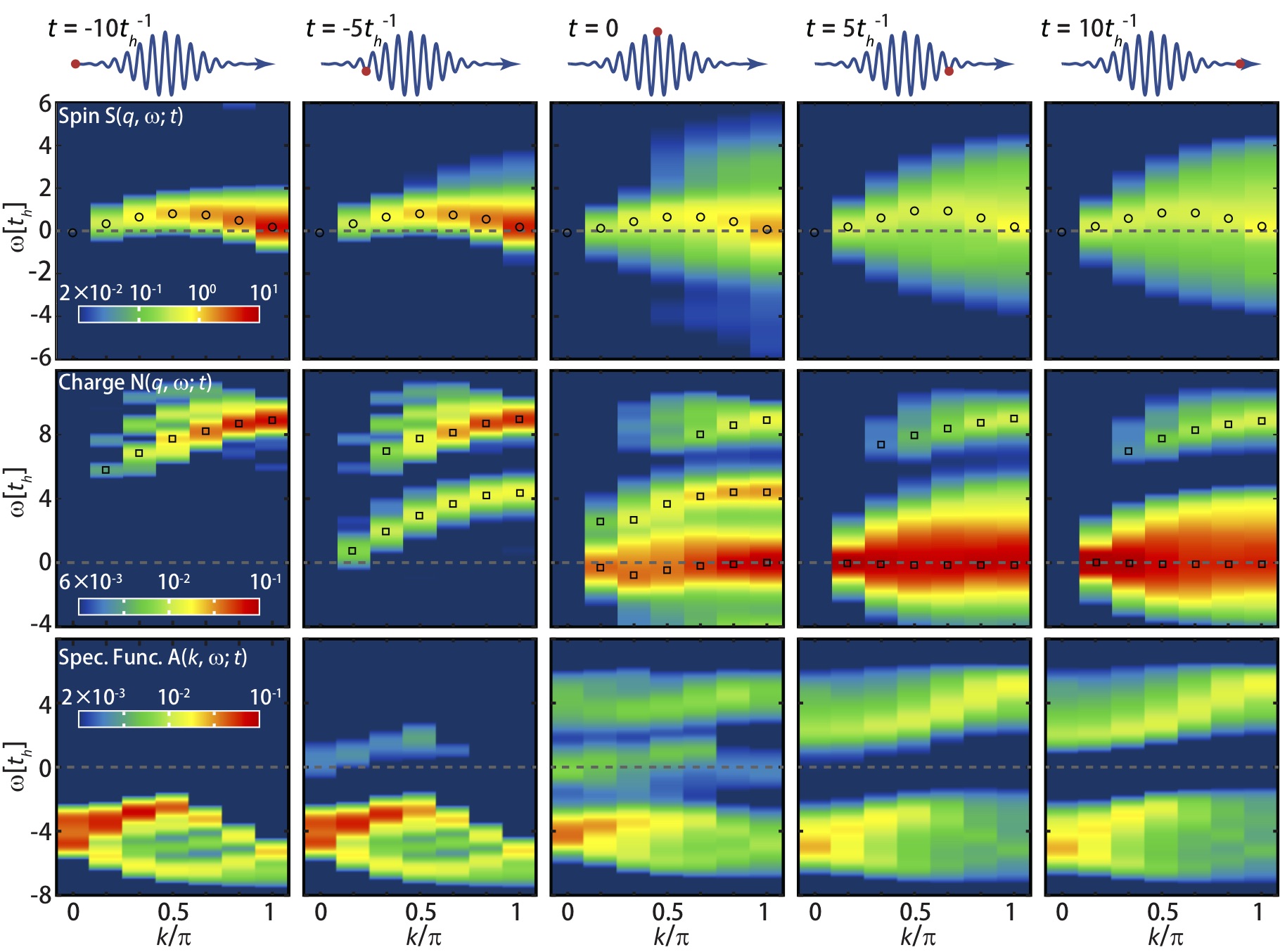}
\caption{\label{fig:2} (Color online) Snapshots of dynamical spin $S(q,\omega;t)$ [upper panels], charge [middle panels] structure factors $N(q,\omega;t)$ and single particle spectral function $A(k,\omega;t)$ [lower panels] during the pump at $t=-10$, $-5$, $0$, $5$ and $10t_h^{-1}$. Those open markers denote the peak positions at various momenta. The top insets indicate the current time (red dot) compared to the pulsed laser. 
}
\end{center}
\end{figure*}
To characterize the evolution of elementary excitations, we consider a nonequilibrium analog of the dynamical spin and charge structure factors. Assuming the probe pulse weak enough ($\Ham_{\rm probe}\!\ll\! \Ham$) to be treated by perturbation and neglecting the matrix-element effect, the time-dependent cross-section can be written as\cite{freericks2009theoretical}
\begin{equation}\label{pumpprobeintensity}
I(\omega,t)\! \propto\!\iint d\tau_1d\tau_2\ e^{i\omega (\tau_1 -\tau_2)} g(\tau_1;t) g(\tau_2; t) C(\tau_1,\tau_2)
\end{equation}
where the $g(\tau;t)$ is a probe shape function at time $t$, taken as Gaussian $\exp[-(\tau-t)^2/2\sigma_{\rm pr}^2]/{\sqrt{2\pi}\sigma_{\rm pr}}$ in this work.
The nonequilibrium correlation function is defined as
$C(\tau_1, \tau_2) = \langle \hat{\mathcal{O}}^\dagger(\tau_1)\hat{\mathcal{O}}(\tau_2)\rangle$ and $\hat{\mathcal{O}}$ is a generic observable for a given measurement. For convenience, the prefactor can be defined as a transformation matrix $f(\tau_1,\tau_2; \omega, t) = e^{i\omega (\tau_1 -\tau_2)} g(\tau_1;t) g(\tau_2; t)$, from a two-time $(\tau_1,\tau_2)$ to a time-frequency $(\omega, t)$ domain.
Therefore, the pump-probe spin/charge cross-section reads as
\begin{equation}\label{pumpprobeintensity}
S(\textbf{q},\omega;t) = \frac1{2\pi}\iint d\tau_1d\tau_2\ f(\tau_1,\tau_2; \omega, t) S_\textbf{q}(\tau_1,\tau_2)
\end{equation}
and the single-particle Green's function
\begin{equation}\label{pumpprobeintensity}
A(\textbf{k},\omega;t) = \frac1{2\pi i}\iint d\tau_1d\tau_2\ f(\tau_1,\tau_2; \omega, t) G^<_\textbf{k}(\tau_1,\tau_2)
\end{equation}
where the non-equilibrium correlation function is defined as
$S_\textbf{q}(\tau_1, \tau_2) = \langle {\rho_{-\textbf{q}}}^\dagger(\tau_1){\rho_\textbf{q}}(\tau_2)\rangle$ and ${\rho_\textbf{q}}$ is charge or spin density operator. 
Due to the uncertainty principle and the necessity to highlight the dynamical properties, we choose a probe width $\sigma_{\rm pr}=2.0t_h^{-1}$ as a compromise between time and frequency resolutions.

\begin{figure*}[!ht]
\begin{center}
\includegraphics[width=16cm]{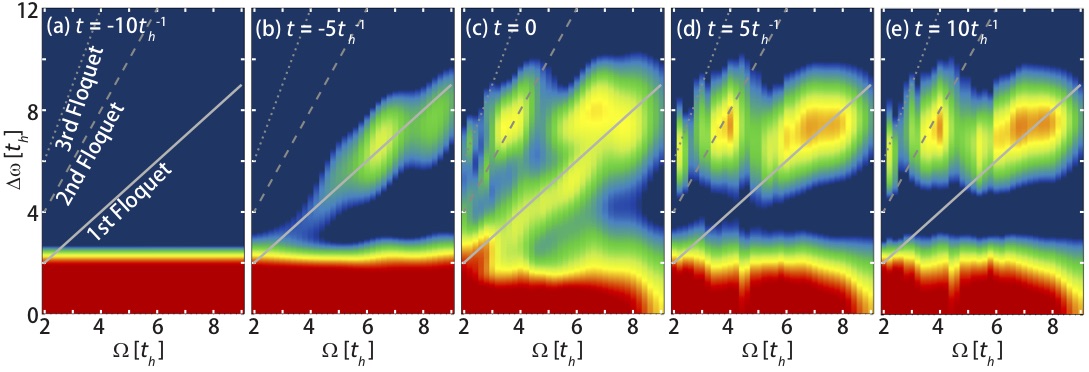}
\caption{\label{fig:3} (Color online) (a-e) Snapshots of autocorrelation $\mathcal{C}(k,\Delta\omega;t)$ for $k=0$. The solid/dashed/dotted lines denote the first/second/third Floquet frequencies as a function of pump. The time frames are the same as those in Fig.~\ref{fig:2}. 
}
\end{center}
\end{figure*}

The upper and middle panels in Fig.~\ref{fig:2} show the (momentum-energy-resolved) dynamical spin and charge structure factors as a function of time, starting from the zero-temperature ground state at $t\!=\!-10t_h^{-1}$. 
Since we are interested in the evolution of different physical processes, let us first focus on the ramp-up regime of pump field ($t\leq 0$).
As shown in the top panels of Fig.~\ref{fig:2}, the spin structure factor $S(q,\omega;t)$ suffers from an overall drop of spectral weight due to the disturbance of the magnetic background, seen in Fig.~\ref{fig:1}.
Although they are rather weak due to the already suppressed spectral weight, we can distinguish the replica bands above and below the spinon excitations close to the peak of the pump envelope ($t= 0$).

At the same time, the charge structure factor $N(q,\omega;t)$ [middle panels of Fig.~\ref{fig:2}] clearly shows a transient, parallel sideband of excitations within the Mott gap. The interval is roughly $\Omega$, hinting at the underlying connection to Floquet theory in the multi-particle channel. 
As time progresses and the field grows in strength, the first sideband gains increasing weight and a second sideband starts to develop and grow. 
Meanwhile, the shapes of both the original and replica progressively flatten, a signature of the bandwidth renormalization, further indicating the dominance of Floquet physics for $t\leq 0$ as discussed in Appendix \ref{app:3_1}.

Due to this coincidence with Floquet theory, we compare the numerically calculated nonequilibrium structure factors with those obtained from analytical, adiabatic Floquet calculations. 
While the charge response qualitatively mirrors the predictions from Floquet theory for an infinitely wide pump profile, the spin response deviates significantly [see the comparison and discussions in Appendix~\ref{app:2}]. 
It suggests that though the Floquet states play an important role in this ramp-up regime, the correct occupancy does not adiabatically follow a pure Floquet state for pumping close to resonance.

The nonequilibrium elementary excitations and their linkage to the transient virtual states can be better illustrated through the evolution of electronic structure. The bottom panels of Fig.~\ref{fig:2} show the time-dependent single-particle spectral function $A(k,\omega;t)$. 
As expected, the transient Floquet sideband and consequent bandwidth renormalization develop at the beginning of pump. 
These sidebands broaden since the accumulation of dressed electrons displays the electron-electron interactions. 
Thus, the scattering across those sidebands then accounts for the appearance of features at integer values of the phonon energy in the multi-particle channel.

The transient dynamics at the beginning can be mostly attributed to virtual states in both multi- and single-particle channel. However, since we are considering an ultrafast pump condition with a finite width, it is natural to question how these excitations project into the remnant states after the pump.
In fact, the nonequilibrium physics with a time profile deviates from the Floquet scenario. This deviation may be higher-order corrections with the increase of field and transient states, but becomes evident with the decrease of the pump (\textit{i.e.} $t>0$ panels in Fig.~\ref{fig:2}).
From the single-particle level, both the lower and upper Hubbard bands tend to recover from virtual states, however, leaving some remnant signatures of pump: the anti-holon tail at $k\!>\!\pi/2$ in the lower Hubbard band indicates the doped mobile holes, while the upper Hubbard band becomes dispersive and maintains spectral weight accounting for the retained populations reflected in Fig.~\ref{fig:1}. 
The low-energy charge excitations persist after the pump,  indicating their association with real rather than virtual electronic states.
These highly gapless modes coexist with the Mott gap (and Mott excitations $\sim U$), reflecting the intrinsically coherent particle-hole excitations induced by the pump.
This population/depletion makes low-energy charge excitations possible with the extra scattering pathways within each band, and also accounts for the suppression of spin order.

\section{Analysis and Discussions}

The above discussions already indicate that the transient nonequilibrium electronic structure and elementary excitations rely on the appearance of virtual sidebands, but their remnant distribution after pump deviates from a simple Floquet picture and tends to be dominated by the many-body interactions.
To understand how these two scenarios are related dynamically during the pump, we further examine the dependence on pump frequency $\Omega$. 
Define the autocorrelation of spectral function
\begin{equation}\label{autocorr}
\mathcal{C}(k,\Delta\omega;t) = \int_{-\infty}^{\infty} A(k,\omega;t) A(k,\omega+\Delta\omega;t)\,d\omega,
\end{equation}
which represents the self-similarity of nonequilibrium spectral function with an energy shift $\Delta\omega$. 
The photon-dressed uncorrelated electrons lead to an autocorrelation peaked at integer values of pump frequencies [indicated by the gray lines in Fig.~\ref{fig:3}], while instantaneously quenched excitations should locate horizontally. 
Therefore, the map of $\mathcal{C}(k,\Delta\omega;t)$ as a function of pump frequency $\Omega$ reflects how the correlated electrons obey or violate both extreme scenarios instantaneously.

Considering higher electron density $n_k$, we focus on examining the autocorrelation at the $\Gamma$ point for the corresponding times and continuously tune the pump frequency $\Omega$ from $2t_h$ (close to resolution limit) to $9t_h$ (beyond resonance) as shown in Fig.~\ref{fig:3}. For a typical choice of $t_h=350$meV in cuprate, this pump frequency lies in the range of $0.7-3.15$eV.
The nontrivial peaks of $\mathcal{C}(k,\Delta\omega;t)$ lie along the first Floquet replica at the beginning of the pump and then increasingly occupy higher sidebands with growing pump strength.
That is exactly what is expected in a Floquet virtual state scenario. 
However, instead of staying at these virtual states, the autocorrelation weight tends to spread out. This reflects that the accumulated, photon-dressed electrons are scattered in the presence of interactions [see Fig.~\ref{fig:3}(c)],
transferring spectral weights from in-gap Floquet side-bands to real many-body states. 
Then, as the pump pulse tails off, those transient side-bands disappear, either leaving part of the spectral weight redistributed into coherent unoccupied states (upper Hubbard band) or dropping back into the original states.

\begin{figure}[!t]
\begin{center}
\includegraphics[width=8.5cm]{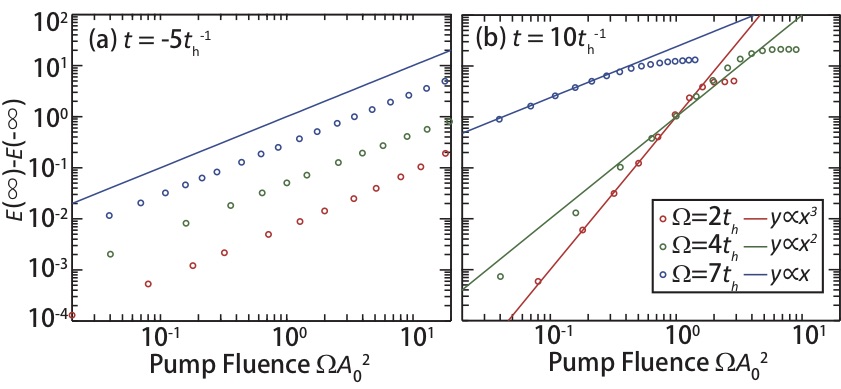}
\caption{\label{fig:4} (Color online) Final photoinjected energies (open circles) (a) at the beginning ($t=-5t_h^{-1}$) and (b) after the pump ($t=10t_h^{-1}$) as a function of pump fluences, at $\Omega=2,\ 4,\ 7t_h$. The colored straight lines represents a first, second and third order polynomial fitting.
}
\end{center}
\end{figure}

This transition is also reflected by the photoinjected energies. As shown in Fig.~\ref{fig:4}, the absorbed energy starts via a linear relation with respect to pump fluences, at which time the dynamics is dominated by Floquet virtual states. However, the final energy changes (below the saturation) scale with the pump fluencies through third-, second- and first-order polynomials at $\Omega = 2$, 4, and 7$t_h$, respectively. 
Therefore, the energy absorptions follow an integer-photon dipole transition picture at the end of the pump.
Compared with the Fig.~\ref{fig:3}(e), these frequencies correspond to the resonance with third, second and first Floquet sidebands. 
Thus, these dynamical spectra link the two extreme scenarios over a pump process.
Although starting from the virtual Floquet states, the single-body electrons are no longer the eigenstates of a correlated system and the impact of the external field coupled to charge scrambles these virtual states.
Toward the end of the pump field, the interaction of dress electrons becomes evident, driving the transition of transient states to remnant excitations at resonance energies and momenta. 
The net post-pump effect is then a selective quench of populations and elementary excitations.

\section{Conclusions}
To summarize, we have studied the time-dependent nonequilibrium and nonsteady state spectroscopies of correlated electrons for a model Hamiltonian coupled to an ultrafast pulsed field. 
The dynamics of spin and charge excitations displays spectral structure associated with Floquet virtual states in the presence of the pump, but evolve into metal-like gapless spectra as the pump field tails off.
We dynamically connect to the effective quantum quench and show that the post-pump remnant excitations develop through a coherent transition --- which can be understood as a resonance between transient Floquet side-bands --- followed by an interaction-induced broadening of the population transfer. 
Photon-dressed electrons are then scattered to the unoccupied upper Hubbard band, stabilizing the low-energy charge and suppressing spin excitations. However, the Mott gap persists, suggesting that these photoinduced excitations are created selectively rather than via incoherent heating at ultrashort time scales.
Thus, we show that these well-known coherent and incoherent approximations provide effective interpretations only at certain time regimes.
This detailed understanding of pathways of photoinduced excitations provides a foundation for Floquet engineering of exotic phenomena (including topological insulators\cite{dehghani2015optical,iadecola2015occupation}, Weyl semimetals\cite{hubener2017creating,zhang2016theory}, superconductors\cite{wang2017light,bittner2017light} and frustrated systems\cite{claassen2017dynamical}) in a post-pump regime.
It is thus possible to selectively quench elementary excitations with a designed laser pulse, and furthermore, to coherently tune the emergent properties of correlated materials through an ultrafast technique.

\section*{ACKNOWLEDGEMENTS}

This work was supported at SLAC and Stanford University by the U.S. Department of Energy, Office of Basic Energy Sciences, Division of Materials Sciences and Engineering, under Contract No. DE-AC02-76SF00515. Y.W. was supported by the Stanford Graduate Fellows in Science and Engineering. A portion of the computational work was performed using the resources of the National Energy Research Scientific Computing Center supported by the U.S. Department of Energy, Office of Science, under Contract No. DE-AC02-05CH11231.

\appendix
\setcounter{equation}{0}
\renewcommand\theequation{A\arabic{equation}}

\section{Comparison with Finite Temperature}\label{app:1}
\begin{figure}[!th]
\begin{center}
\includegraphics[width=8cm]{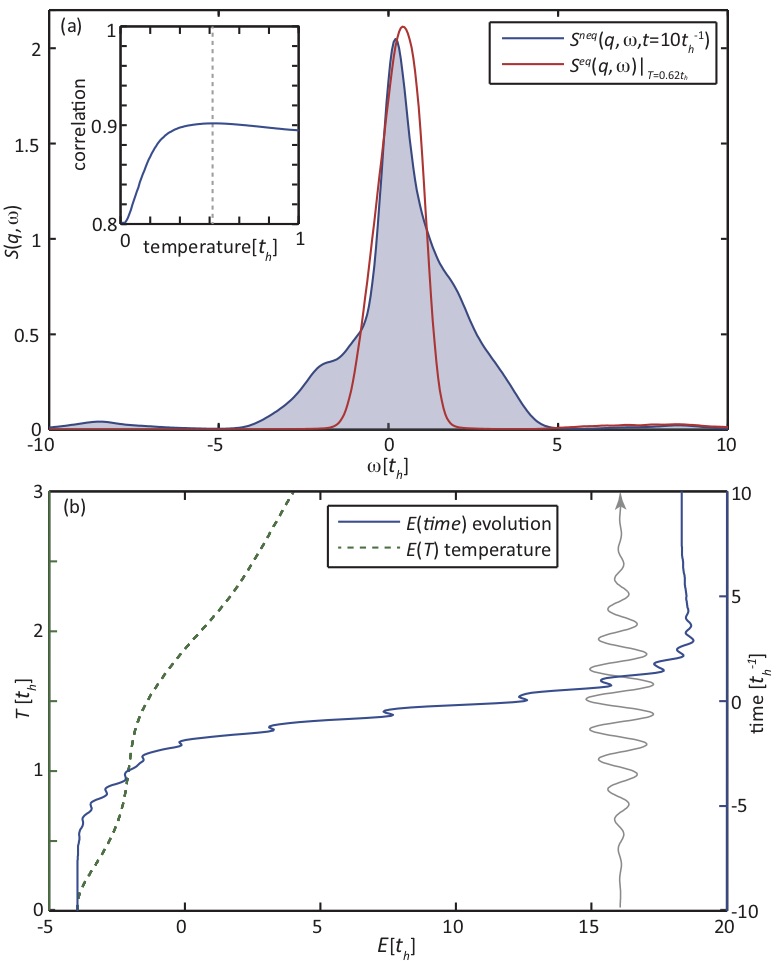}
\caption{\label{fig:comp} (a) Comparison of finite temperature spectra $S^{eq}(\pi,\omega)|_T$ (red) with the nonequilibrium $S^{neq}(\pi,\omega;t)$ (blue shaded region) at $t=10t_h^{-1}$ after pump. The effective temperature is determined through the maximum correlation principle shown in the inset. 
(b) Evolution of average energy $E(t)$ (blue solid line) and temperature dependence of equilibrium energy $E(T)$ (green dashed line). The pump pulse in both figures is set as Fig.~\ref{fig:2}.
}
\end{center}
\end{figure}

To show the pumped elementary excitations are not simply heating effects, we compare the post-pump spectroscopies and energy with those evaluated at a finite temperature. Taking the most relevant spin dynamical structure factor $S(q,\omega)$ at $q=\pi$ as an example, the effective temperatures extracted from both approaches do not match and cannot explain the spectral shape.

Fig.~\ref{fig:comp}(a) shows the determination of effective temperature from the similarity of spectra. To distinguish the dynamical spin structure factor obtained from different origins, we denote the nonequilibrium one (shown in the main text) as $S^{\rm neq}(q,\omega;t)$ while the finite-temperature equilibrium one as $S^{eq}(q,\omega)|_{T}$ with $T$ standing for the temperature. The similarity of spectra is reflected by the correlation
\begin{equation}\label{corr}
\textrm{corr}(T,t) = \frac1{\mathcal{N}}\int_{-\infty}^\infty S^{\rm neq}(q,\omega;t)\, S^{eq}(q,\omega)|_{T}\, d\omega
\end{equation}
where $\mathcal{N}$ stands for the normalization factor $\left[\int_{-\infty}^\infty \left[S^{eq}(q,\omega)|_{T}\right]^2d\omega \int_{-\infty}^\infty \left[S^{\rm neq}(q,\omega;t)\right]^2 d\omega \right]^{1/2}$. Thus, the correlation is maximized as 1 at $T=0$ and $t=-\infty$. Concerning the heating effect after pump, the inset of Fig.~\ref{fig:comp}(a) shows the correlation between post-pump spectra $S^{\rm neq}(q,\omega;t=10t_h^{-1})$ and various finite temperatures, indicating an ``effective temperature'' of $T_{\rm eff}=0.62t_h$. However, the equilibrium $S(q,\omega)$ at $T_{\rm eff}$ (red curve) cannot reflect the features of pumped spin structure factor (blue curve). 

On the other hand, Fig.~\ref{fig:comp}(b) shows the comparison of nonequilibrium average energy $E(t)$ (blue solid line) and finite-temperature energy $E(T)$ (green dashed line). Obviously, the effective temperature extracted from energy indicated a $T_{\rm eff} \gg 3t_h$, far beyond the estimation of the spectrum similarity. In fact, considering the eigenspectrum radius is $~UN/2\approx 48t_h$, the energy in pumped system $E(t)$ almost reaches infinite temperature, if it is a heating effect.

The disagreement of both approaches reflects the pumped correlated electrons at least at moderate intensity creates coherent excitations instead of heating.

\setcounter{equation}{0}
\renewcommand\theequation{B\arabic{equation}}
\section{Comparison with Floquet Theory}\label{app:2}
\subsection{A Brief Introduction to the Floquet Theory}
\begin{figure*}[!ht]
\begin{center}
\includegraphics[width=15cm]{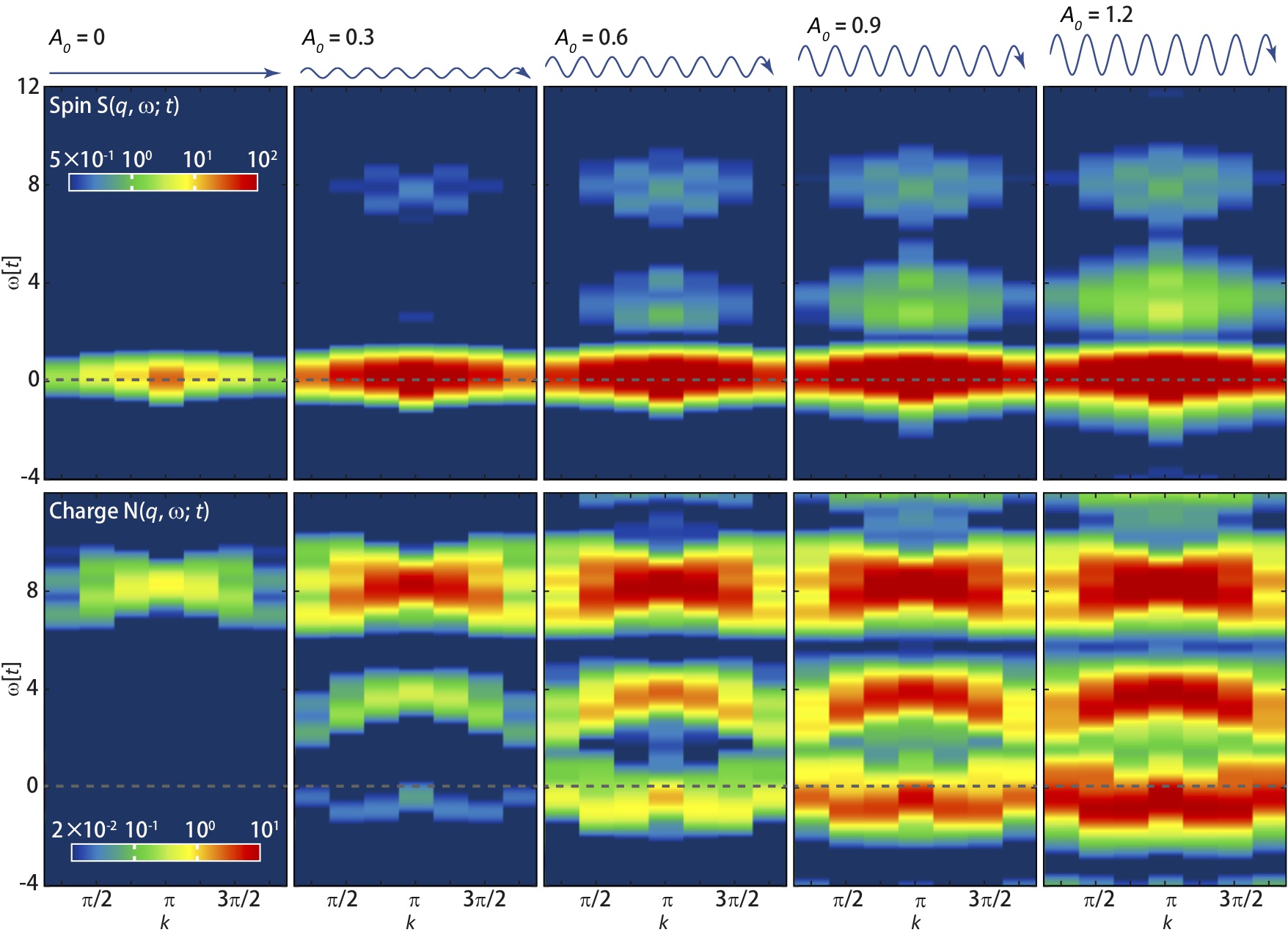}
\caption{\label{fig:adFloquet} The spin (upper panels) and charge (lower panels) structure factors evaluated by adiabatic Floquet approximation at an eight-site chain. The top insets indicate the steady pump field to mimic the time evolution.
}
\end{center}
\end{figure*}

Due to the connection of the pumped charge and spin excitations to the Floquet sidebands, we calculate the structure factors obtained from Floquet theory with various pump strength. In the Bloch-Floquet framework, the \Schrdg\ equation 
\begin{eqnarray}\label{FloquetSchrodinger}
i\frac{\partial }{\partial t} |\Psi(t)\rangle = \Ham\big[\bar{A}(t)\big] |\Psi(t)\rangle
\end{eqnarray}
with periodic pump field $\bar{A}(t+2\pi/\Omega) = \bar{A}(t)$ has the fundamental solutions
\begin{eqnarray}\label{FloquetSolution}
|\Psi_\lambda(t)\rangle=e^{-i\epsilon_\lambda t}\sum_{\alpha m} u^{(\lambda)}_{m\alpha} |m,\alpha\rangle.
\end{eqnarray} 
in a direct product basis $|m; \alpha\rangle = |m\rangle \otimes |\alpha\rangle$ with $\langle t|m\rangle = e^{-im\Omega t}$. Then, the problem is equivalent to solving a Floquet Hamiltonian
\begin{eqnarray}\label{FloquetHam}
\Ham_F &=& \sum_{mm^\prime} \left[ H_{m^\prime-m} -m\Omega\delta_{mm^\prime} \right] |m^\prime\rangle \langle m|\nonumber\\
&=& \left(\begin{array}{ccccc}
          \ddots & \ddots & \vdots & \vdots &  \\
          \ddots & H_0+\Omega & H_{-1} & H_{-2} & \cdots\\
          \cdots & H_1 & H_0 & H_{-1} & \cdots \\
          \cdots & H_2 & H_1 & H_0-\Omega & \ddots \\
           & \vdots & \vdots & \ddots & \ddots
        \end{array}
\right).
\end{eqnarray}
with $\Ham(t)=\sum_m H_m |m\rangle$. The choice of coefficients $u^{(\lambda)}_{m\alpha}$ defines a unique wavefunction among the infinite fundamental solutions. 

\subsection{Adiabatic Floquet Evaluation of Charge and Spin Structure Factors}
Under realistic pump condition, both $A(t)$ and $\Ham(t)$ are not strictly periodic. In this case, we consider the adiabatic approximation to determine what $u^{(\lambda)}_{m\alpha}$ coefficients correspond to the instantaneous state at a given time $t$. That is, the realistic wavefunction at $t$ is mimicked by a Floquet solution of periodic Hamiltonian $\bar{\Ham}[\bar{A}_0\cos(\Omega t)]$, where the amplitude is chosen to be the realistic instantaneous pump strength $\bar{A}_0 = \bar{A}(t)$. 
The Floquet coefficients $u^{(\lambda)}_{m\alpha}$ at time $t$ are then determined by the maximum overlap with the wavefunction of $t-\delta t$.

Fig.~\ref{fig:adFloquet} shows the adiabatic evaluation of spin and charge structure factors at an increasing pump field. Compared to the first half of pumped spectra in Fig.~\ref{fig:2}, the charge structure factor $N(q,\omega;t)$ qualitatively matches the features of low-energy side bands; however, the $S(q,\omega;t)$ displays more of higher-energy excitations than spectral weight suppression, opposed to the ED calculations. 
The cause is two-fold: First, in our model photons only couple to charge, therefore, any Floquet modification of the underlying spin dynamics is necessarily a higher-order effect. Second, the spin excitation spectrum is gapless already in equilibrium, hence there is no well-defined notion of adiabatic continuity to a transient Floquet analog. Deviations from an adiabatic transient can be expected to be most evident in the spin channel.

\subsection{Bandwidth Renormalization}\label{app:3_1}
Consider a steady-state scenario with vector potential $A(t) = A_0\cos(\Omega t +\phi)$ upon a tight-binding model $\Ham_0 = -2t\sum_{k\sigma}\cos k\, n_{k\sigma}$ through Peierls substitution
\begin{eqnarray}
\Ham(t) = -2t_h\sum_{k\sigma}\cos \Big(k-A(t)\Big) n_{k\sigma}
\end{eqnarray}
Then, the blocks in Eq.~\ref{FloquetHam} read as
\begin{eqnarray}
H_m &=& \frac{-2t_h}{T} \sum_{k\sigma}\int_0^Tdt \cos \Big(k-A_0\cos(\Omega t +\phi)\Big) n_{k\sigma} \nonumber\\
&=& -t_he^{-im\phi}i^m \sum_{k\sigma}\left[ e^{-ik} \mathcal{J}_m(A_0) + e^{ik} \mathcal{J}_m(-A_0) \right] n_{k\sigma}\nonumber\\
\end{eqnarray}
where $\mathcal{J}_m(z)$ is the Bessel function of the first kind. 
For small $A_0$ there is $\mathcal{J}_{m\neq0}(A_0) \ll \mathcal{J}_0(A_0)$, therefore, only the diagonal blocks $-2t_h\mathcal{J}_0(A_0)\sum_{k\sigma}\cos(k)n_{k\sigma}$ significantly contribute. The effective bandwidth is changed from $4t_h$ to $4t_h\mathcal{J}_0(A_0)$. Since $|\mathcal{J}_0(A_0)|<1$, this always leads to a bandwidth renormalization or shrinkage.

\begin{figure}[!t]
\begin{center}
\includegraphics[width=7cm]{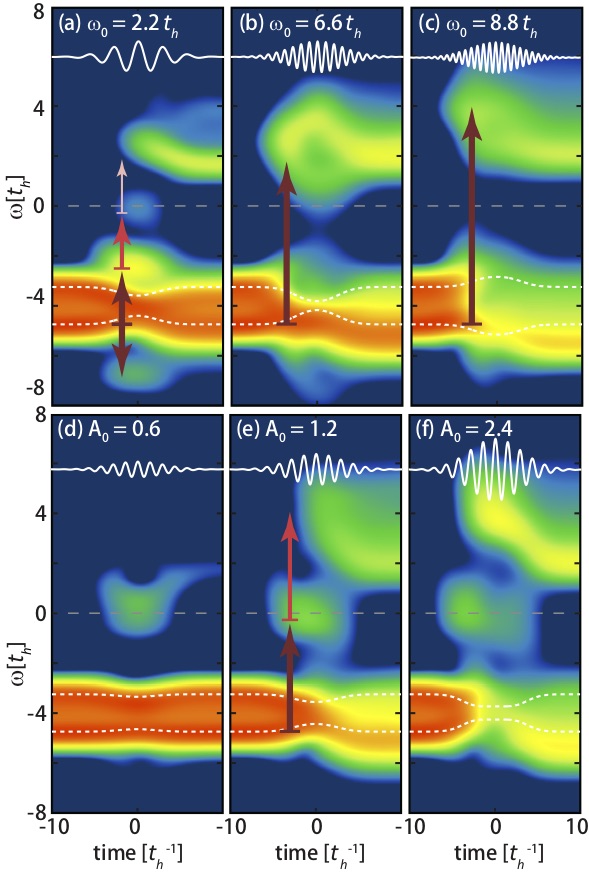}
\caption{\label{fig:S4} (a-c) Evolution of $A(0,\omega)$ for various pump frequency $\omega_0$ with a fixed pump amplitude ($A_0 = 1.2$).
(d-f) Evolution of $A(0,\omega)$ for various pump amplitude $A_0$ with a fixed pump frequency ($\omega_0 = 4.4t_h$). The pump width is fixed for all these figures ($\sigma_t = 3.0t_h^{-1}$). The red arrows guide the eye for the Floquet frequencies, while the dotted white lines denote the Floquet estimated spinon and holon branches. 
}
\end{center}
\end{figure}

\setcounter{equation}{0}
\renewcommand\theequation{C\arabic{equation}}
\section{Continuous Time View}\label{app:3}
\subsection{Amplitude and Frequency Dependence }

In complementary to Figs.~\ref{fig:3} and \ref{fig:4}, we further compare the pump probe spectroscopies as a function of time for various frequencies and strengths to better understand the development of Floquet replicas and their interplay with the closure of pump pulse. Figs.~\ref{fig:S4}(a-c) and (e) show the evolution of $A(0,\omega;t)$ where the spectral weight is clearest for various pump frequencies. The energy positions where discrete spectral weights develop roughly match the Floquet side-bands, in spite of some off-sets due to the finite pocket as well as correlations.  
 By comparison of these different frequencies, one would notice the virtual Floquet bands in the upper Hubbard band restructures and scatters towards the real single-particle dispersion with the decrease or closure of the pump pulse. In contrast, those virtual spectral weights inside the gap or off-resonance have to disappear without any accommodation afterwards [see Fig.~\ref{fig:S4}(a)].

On the other hand, the change of pump amplitude tunes the bandwidth and spectral weight in both upper and lower Hubbard bands, although it shines little influence of the resonance. As shown in Figs.~\ref{fig:S4}(d-f), the spectral weight into the second Floquet side band is usually much weaker than the first at the beginning of pump. Due to the resonance with upper Hubbard bands, however, electrons in the second side band are well accommodated by scattering into the upper Hubbard band. Considering the bandwidth of $m$th side band roughly $\propto \mathcal{J}_m(A_0)$ which drops rapidly with $m$ for small $A_0$, the noticeable occupation of upper Hubbard band only appears when the pump is strong enough. That explains the fact that weak pump [$A_0=0.6$, Fig.~\ref{fig:S4}(d)] excites only the first side band which is off-resonance and disappears when the pump is off.

\subsection{Bandwidth Comparison with Effective $t-J$ Model}
The bandwidth effect can also be reflected by the suppression of lower Hubbard band. We calculate the effective $t-J$ model obtained by Floquet theory at infinite wide pump via $\tilde{t} = \mathcal{J}_0(A)t$ and $\tilde{J} = (4t^2/U) \sum_n \mathcal{J}_{|n|}(A)^2/[1+n\omega_0/U]$ as an estimation for the lower Hubbard band. \cite{mentink2015ultrafast} Accordingly, we evaluate the corresponding holon/spinon branches in this instantaneous effective model at half-filling, indicated by the dotted lines \cite{suzuura1997spin, brunner2000single}. The band renormalizations are consistent with the $t-J$ estimation at the beginning of the pump: the splitting ($\sim 2t_h-J$) is increasingly suppressed with increasing pump strength. This situation becomes different and the $t-J$ estimation no longer works when much spectral weight is drawn out of the lower-Hubbard band near the pulse peak (especially for strong pump), where electrons behave more like Fermi liquid instead of half-filled Mott insulator.

\bibliography{paper}

\end{document}